\documentclass[onecolumn]{aastex63}
\usepackage{xcolor}

\usepackage{natbib}
\usepackage{amsmath}
\usepackage{enumitem}

\shortauthors{Bernardinelli et al.}

\defcitealias{Bernstein2000}{BK}

\newcommand{\E}{\mathrm{E}}

\newcommand{\des}{\textit{DES}}

\newcommand{\eqq}[1]{Equation~(\ref{#1})}

\begin{document}

\title{Testing the isotropy of the Dark Energy Survey's extreme trans-Neptunian objects}

\author[0000-0003-0743-9422]{Pedro H. Bernardinelli}
\affiliation{Department of Physics and Astronomy, University of Pennsylvania, Philadelphia, PA 19104, USA}
\correspondingauthor{Pedro H. Bernardinelli, \url{pedrobe@sas.upenn.edu} \& Stephanie Hamilton, \url{sjhamil@umich.edu}}

\author[0000-0002-8613-8259]{Gary M. Bernstein}
\affiliation{Department of Physics and Astronomy, University of Pennsylvania, Philadelphia, PA 19104, USA}

\author[0000-0003-2764-7093]{Masao Sako}
\affiliation{Department of Physics and Astronomy, University of Pennsylvania, Philadelphia, PA 19104, USA}

\author{Stephanie Hamilton}
\affiliation{Department of Physics, University of Michigan, Ann Arbor, MI 48109, USA}

\author[0000-0001-6942-2736]{David W. Gerdes}
\affiliation{Department of Astronomy, University of Michigan, Ann Arbor, MI 48109, USA}
\affiliation{Department of Physics, University of Michigan, Ann Arbor, MI 48109, USA}

\author[0000-0002-8167-1767]{Fred C. Adams}
\affiliation{Department of Astronomy, University of Michigan, Ann Arbor, MI 48109, USA}
\affiliation{Department of Physics, University of Michigan, Ann Arbor, MI 48109, USA}

\author[0000-0002-8737-742X]{William R. Saunders}
\affiliation{Department of Physics and Astronomy, University of Pennsylvania, Philadelphia, PA 19104, USA}
\affiliation{Department of Astronomy, Boston University, Boston, MA 02215, USA}

\author{M.~Aguena}
\affiliation{Departamento de F\'isica Matem\'atica, Instituto de F\'isica, Universidade de S\~ao Paulo, CP 66318, S\~ao Paulo, SP, 05314-970, Brazil}
\affiliation{Laborat\'orio Interinstitucional de e-Astronomia - LIneA, Rua Gal. Jos\'e Cristino 77, Rio de Janeiro, RJ - 20921-400, Brazil}
\author{S.~Allam}
\affiliation{Fermi National Accelerator Laboratory, P. O. Box 500, Batavia, IL 60510, USA}
\author{S.~Avila}
\affiliation{Instituto de Fisica Teorica UAM/CSIC, Universidad Autonoma de Madrid, 28049 Madrid, Spain}
\author{D.~Brooks}
\affiliation{Department of Physics \& Astronomy, University College London, Gower Street, London, WC1E 6BT, UK}
\author{H.~T.~Diehl}
\affiliation{Fermi National Accelerator Laboratory, P. O. Box 500, Batavia, IL 60510, USA}
\author{P.~Doel}
\affiliation{Department of Physics \& Astronomy, University College London, Gower Street, London, WC1E 6BT, UK}
\author{S.~Everett}
\affiliation{Santa Cruz Institute for Particle Physics, Santa Cruz, CA 95064, USA}
\author{J.~Garc\'ia-Bellido}
\affiliation{Instituto de Fisica Teorica UAM/CSIC, Universidad Autonoma de Madrid, 28049 Madrid, Spain}
\author{E.~Gaztanaga}
\affiliation{Institut d'Estudis Espacials de Catalunya (IEEC), 08034 Barcelona, Spain}
\affiliation{Institute of Space Sciences (ICE, CSIC),  Campus UAB, Carrer de Can Magrans, s/n,  08193 Barcelona, Spain}
\author{R.~A.~Gruendl}
\affiliation{Department of Astronomy, University of Illinois at Urbana-Champaign, 1002 W. Green Street, Urbana, IL 61801, USA}
\affiliation{National Center for Supercomputing Applications, 1205 West Clark St., Urbana, IL 61801, USA}
\author{K.~Honscheid}
\affiliation{Center for Cosmology and Astro-Particle Physics, The Ohio State University, Columbus, OH 43210, USA}
\affiliation{Department of Physics, The Ohio State University, Columbus, OH 43210, USA}
\author{R.~L.~C.~Ogando}
\affiliation{Laborat\'orio Interinstitucional de e-Astronomia - LIneA, Rua Gal. Jos\'e Cristino 77, Rio de Janeiro, RJ - 20921-400, Brazil}
\affiliation{Observat\'orio Nacional, Rua Gal. Jos\'e Cristino 77, Rio de Janeiro, RJ - 20921-400, Brazil}
\author{A.~Palmese}
\affiliation{Fermi National Accelerator Laboratory, P. O. Box 500, Batavia, IL 60510, USA}
\affiliation{Kavli Institute for Cosmological Physics, University of Chicago, Chicago, IL 60637, USA}
\author{D.~L.~Tucker}
\affiliation{Fermi National Accelerator Laboratory, P. O. Box 500, Batavia, IL 60510, USA}
\author{A.~R.~Walker}
\affiliation{Cerro Tololo Inter-American Observatory, NSF’s National Optical-Infrared Astronomy Research Laboratory, Casilla 603, La Serena, Chile}
\author{W.~Wester}
\affiliation{Fermi National Accelerator Laboratory, P. O. Box 500, Batavia, IL 60510, USA}
\collaboration{1000}{(The \des\ Collaboration)}

\begin{abstract}
We test whether the population of ``extreme'' trans-Neptunian objects (eTNOs) detected in the Y4 Dark Energy Survey (\des) data exhibit azimuthal asymmetries which might be evidence of gravitational perturbations from an unseen super-Earth in a distant orbit.  By rotating the orbits of the detected eTNOs, we construct a synthetic population which, when subject to the \des\ selection function, reproduces the detected distribution of eTNOs in the orbital elements $a,e,$ and $i$ as well as absolute magnitude $H$, but has uniform distributions in mean anomaly $\mathcal{M}$, longitude of ascending node $\Omega,$ and argument of perihelion $\omega.$  We then compare the detected distributions in each of $\Omega, \omega,$ and longitude of perihelion $\varpi\equiv\Omega+\omega$ to those expected from the isotropic population, using Kuiper's variant of the Kolmogorov-Smirnov test.  The three angles are tested for each of 4 definitions of the eTNO population, choosing among $a>(150,250)$~AU and perihelion $q>(30,37)$~AU.  These choices yield 3--7 eTNOs in the \des\ Y4 sample.  Among the twelve total tests, two have the likelihood of drawing the observed angles from the isotropic population at $p<0.03.$  The 3 detections at $a>250, q>37$~AU, and the 4 detections at $a>250, q>30$~AU, have $\Omega$ distribution with $p\approx0.03$ of coming from the isotropic construction, but this is not strong evidence of anisotropy given the 12 different tests.  The \des\ data taken on their own are thus consistent with azimuthal isotropy and do not require a ``Planet 9'' hypothesis. The limited sky coverage and object count mean, however, that the \des\ data by no means falsify this hypothesis.
\vspace{0.2in}
\end{abstract}
\reportnum{DES-2019-0512}
\reportnum{FERMILAB-PUB-20-113-AE}

\section{Introduction} \label{sec:intro}
\citet{Trujillo2014} noted that the sample of then-known trans-Neptunian objects (TNOs) with semi-major axis $a > 150$ AU and perihelion $q > 30$ AU seemed clustered in their arguments of perihelion near $\omega \approx 0\degr$. \citet{Batygin2016} argue that TNOs with $a > 250$ AU are also clustered in their longitude of ascending node, at $90\degr \lesssim \Omega \lesssim 180\degr$, defining the direction of the orbital pole.  They also find clustering in longitude of perihelion, at $ 0\degr \lesssim \varpi \equiv \Omega + \omega \lesssim 90 \degr$ (the apsidal orientation of the orbit), which would indicate a physical alignment of the orbits. The hypothesized dynamical mechanism to stabilize these angles is the presence of a distant planetary-mass perturber (``Planet 9''), extensively reviewed in \cite{Batygin2019}, but question remains as to the statistical significance of this clustering in the face of survey selection effects \citep{Shankman2017,Lawler2016,Kavelaars2019}. The proposed perturber can also generate high-inclination orbits \citep{Batygin2016b,Batygin2017}, and in some scenarios account for the obliquity of the Sun \citep{Bailey2016,Gomes2016,Lai2016}. The inclination instability mechanism proposed in \citet{Madigan2016} can also potentially account for both the argument of perihelion and apsidal clustering \citep{Zderic2020} without a ninth planet. 

Since \citet{Trujillo2014}, numerous other of these ``extreme'' TNOs (eTNOs; $a > 150 \, \mathrm{AU},$ $q > 30 \, \mathrm{AU}$) have been discovered \citep{Bannister2016,SheppardTrujillo2016,Bannister2018,Becker2018,Khain2018,Khain2020,2019AJ....157..139S,Bernardinelli2019}. \cite{Shankman2017} present an analysis of the OSSOS \citep{Bannister2016,Bannister2018} sample of extreme TNOs, using a survey simulator to demonstrate the non-intuitive biases involved in detecting such objects, and to conclude that the distribution of the 8 OSSOS eTNOs is consistent with uniformity in $\Omega, \omega,$ and $\varpi.$\footnote{\cite{Bannister2018} repeat the test with one more object.}
\cite{2019AJ....157..139S} find a modest-significance clustering in the objects with low observational biases, and the analysis of the Minor Planet Center sample by \cite{Batygin2016}, \cite{Brown2017} and \cite{Brown2019} find that there is a small chance of accidental clustering of these objects, albeit with less complete information about the selection function of the discovery surveys. \cite{Trujillo2020} reviews the observational evidence and the statistical significance of the alignment in the distant TNO populations.

We conduct here an independent test of azimuthal isotropy using the eTNOs detected by the \emph{Dark Energy Survey} \citep[\des,][]{Bernardinelli2019}, fully accounting for this survey's observational characteristics and recoverability.  More precisely: we seek a model of the underlying population of eTNOs which (1) is uniformly distributed in $\Omega$ and $\omega$ (and hence in $\varpi$) as well as in mean anomaly $\mathcal{M}$, and which (2) after applying the survey selection function, predicts a distribution in $\{a,e,i,H,\Omega,\omega\}$ which is consistent with that of the true eTNO sample.  If we find such an isotropic distribution which matches the observations, we cannot claim evidence of orbital alignments in the \des\ Y4 eTNO sample. A similar analysis using this survey's difference imaging sample has been presented in \cite{Hamilton2019} and is summarized in Section~\ref{sec:stephanie}.

\section{Sample of extreme trans-Neptunian objects}
\label{sec:sample}

The \des\ surveyed 5000 $\mathrm{deg}^2$ of sky repeatedly over six observing seasons (2013--2019) with the 3 $\mathrm{deg}^2$, 520 Mpix Dark Energy Camera \citep[][DECam]{Flaugher2015} in the $grizY$ optical/NIR bands. The full (wide) survey tiles the footprint with $10\times 90$ seconds exposures in the $griz$ bands and $6 \times 45 + 2 \times 90$ seconds exposures in $Y$ band, with a total of $\approx 80,000$ exposures. \citet{Bernardinelli2019} describe the methodology that allows the recovery of TNOs in the \des, and present a catalog of 316 TNOs detected in the first four years of the survey (Y4; $ \approx 60,000$ images), with typical $r$ band exposures being complete to $r \sim 23.5$. These objects have multi-year arcs, at least 6 unique nights of detections, and $grizY$ photometry, yielding uncertainties in orbital elements and $H$  that are negligible for the isotropy test ($\frac{\sigma_a}{a} \lesssim 3 \% $, $\sigma \lesssim 0.5\degr$ for all angular variables, and $\sigma_H \lesssim 0.1$ mag). 
\citet{Bernardinelli2019} also introduce a methodology for testing the completeness of the survey, which will be extended in this work.

\begin{figure}[tb]
	\centering
	\includegraphics[width=\textwidth]{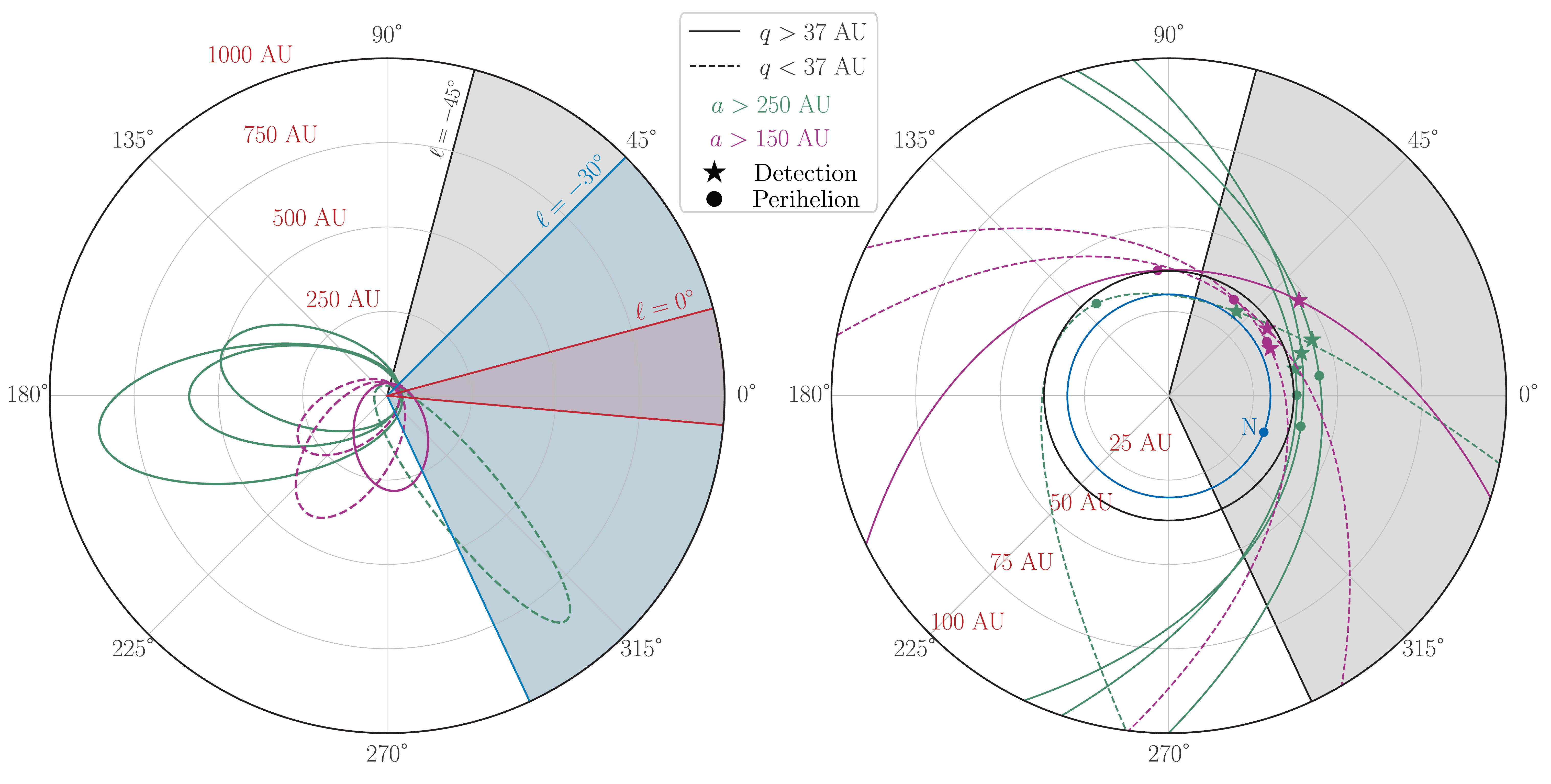}

	\caption{Ecliptic $xy$ plane projection of the orbits of the seven eTNOs. The gray shaded sector in both panels represents the ecliptic extent of the \des\ footprint at ecliptic latitude $\ell = -45\degr$, and the longitudinal extent of the footprint at lower $\ell$ is denoted by the red ($\ell = 0\degr$) and blue ($\ell = -30\degr$) radial lines/sectors in the left panel. Solid lines represent orbits with $q > 37$ AU, while dashed represent $30<q < 37$ AU. The green lines correspond to objects with $a > 250$ AU, and the purple to the ones with $150 < a < 250$ AU.  The right panel presents a closer view of the orbits, with a star denoting the location of each object at the time of its detection and circles marking their perihelia. The blue circle marks Neptune's orbit (and the blue dot its location at barycentric Julian date 2016.0), and the black one represents a distance of 37 AU from the center. \label{im:orbits}}
\end{figure}

Among the 316 objects of the Y4 sample, seven satisfy the original eTNO definition of \citet{Trujillo2014}: $a > 150 \, \mathrm{AU}$ and $q > 30 \, \mathrm{AU}.$ The barycentric orbital elements and absolute magnitudes of these objects are presented in Table \ref{tb:elements}. The ecliptic-plane projection of the orbits, as well as a projection of \des's footprint, are plotted in Figure~\ref{im:orbits}. We refer the reader to Figures 1 and 19 of \cite{Bernardinelli2019} for images of the full \des\ footprint.  Given that the angular clustering  in $\{ \Omega, \omega, \varpi \}$ has been claimed to be present in a variety of subsets of this loosest definition, we will conduct our tests for four cases:
\begin{enumerate}
	\item $a>150$~AU, $q>30$~AU (the full 7-object set), as in \cite{Trujillo2014},
	\item $a>250$~AU, $q>30$~AU (4 \des\ objects), where \cite{Batygin2016} find there is a clustering in $\Omega$ and $\varpi$;
	\item $a>150$~AU,  $q>37$~AU (4 \des\ objects), eliminating objects with the stronger interactions with Neptune (\citealp{Lykawka2007}; see also discussion on \citealp{Shankman2017});
	\item $a>250$~AU, $q > 37$~AU (3 \des\ objects), combining both restrictions. 
\end{enumerate}
The objects belonging to the fourth case are the ones least influenced by Neptune and thus offer the cleanest test for influences from a Planet 9.  Given the small observed population for case (4), however, the tests are going to be weak, and we are wise to also examine the less-restrictive cases (1)--(3) despite potentially weaker signals.

We note that the \des\ eTNO sample has no overlap with the objects analyzed by \cite{Batygin2016}, \cite{Brown2017}, \cite{2019AJ....157..139S}, nor with the OSSOS sample of \cite{Shankman2017}, thus making this test largely statistically independent of these predecessors. Despite this independence, the distributions of $\omega$, $\Omega$ and $\varpi$ for the DES sample (see Table \ref{tb:elements}) show tendencies to lie in the ranges earlier suggested as being over-populated.  It is of interest, therefore, to see if the apparent clustering in this independent sample can be explained as a selection effect.

\begin{deluxetable}{|c | r@{.}l r@{.}l r@{.}l r@{.}l r@{.}l r@{.}l r@{.}l  cc | c |}
  \tablecaption{Barycentric orbital elements at barycentric Julian date 2016.0, absolute $r$ band magnitude, and $r$ band magnitude at discovery for the sample of eTNOs. See \citet{Bernardinelli2019} for more details and state vectors with full covariance matrices for these objects. The last column indicates which of the cases defined in section \ref{sec:sample} each object belongs to. Objects marked with $\star$ were also used in the \citet{Hamilton2019} analysis (section \ref{sec:stephanie}).
Uncertainties are given in parentheses when they exceed the printed precision (rigorous uncertainties are not available for 2013~RF$_{98}$, orbital elements obtained using JPL Horizons).
\label{tb:elements}}
\tablehead{MPC id
  & \twocolhead{$a$ (AU)}
  & \twocolhead{ $e$ }
  & \twocolhead{$i$ (deg)}
  & \twocolhead{$\Omega$ (deg)}
  & \twocolhead{$\omega$ (deg)}
  & \twocolhead{$\varpi$ (deg)} 
  & \twocolhead{$q$ (AU)} & $H_r$ & $m_r$ & Cases }
	\startdata
	2013 RA$_{109}$ & 463&3(2) & 0&901 & 12&39 & 104&79 & 262&91(1) & 7&71 & 46&0 & 5.9 & 22.6 & 1,2,3,4,$\star$ \\
	2015 BP$_{519}$ & 449&3(8) & 0&922 & 54&11 & 135&21 & 348&06 & 123&27 & 35&2(1) & 4.3 & 21.7 & 1,2,$\star$ \\
	2013 SL$_{102}$ & 314&3(1) & 0&879 & 6&50 & 94&73 & 265&49 & 0&22 & 38&1 & 7.1 & 22.9 & 1,2,3,4,$\star$ \\
	2014 WB$_{556}$ & 289&3(6.2) & 0&853(3) & 24&15  & 114&89 & 234&53(49) & -10&56 & 42&5(1.9) & 7.2 & 23.7 & 1,2,3,4 \\
	2016 SG$_{58}$ & 233&0(1) & 0&849 & 13&22 & 118&97 & 296&29 & 55&27 & 35&1 & 7.2 & 22.8 & 1\\
	2016 QV$_{89}$ & 171&6(2) & 0&767 & 21&38 & 173&21 & 281&08(1) & 94&29 & 40&0(1) & 5.9 & 22.8 & 1,3 \\
	508338 (2015 SO$_{20}$) & 164&7 & 0&799 & 23&41 & 33&63 & 354&78(3) & 28&42 & 33&2 & 6.6 & 21.8 & 1 \\
	\hline 
	2013 RF$_{98}$ & 358&2 & 0&90 & 23&54 & 67&63 & 312&05 & 19&68 & 36&1 & 8.6 & 24.2 & $\star$
	\enddata
\end{deluxetable}

\section{Simulated isotropic population}
\label{sec:sim}
We use a simple construction to create a population that is isotropic in $\{\mathcal{M},\Omega,\omega\}$ but predicts a distribution $p(a,e,i,H | s)$ (conditioned on successful detection $s$) that is consistent with that of the detected eTNOs.  Indexing the latter by $j$, we posit an underlying population with
\begin{equation}
  p(a,e,i,\Omega,\omega,\mathcal{M}, H) \propto \sum_j \frac{\delta(a-a_j) \delta(e-e_j) \delta(i-i_j) \delta(H-H_j)}{p(s | a_j, e_j, i_j, H_j)} u(\mathcal{M}) u(\Omega) u(\omega).
  \label{eq:parent}
\end{equation}
In this equation, $u(\theta)$ is a uniform distribution over $\theta \in [0,2\pi]$, $\delta$ is the Dirac delta function, and $p(s | a,e,i,H)$ is the probability of detection of an eTNO in \des\ when averaged over $(\mathcal{M},\Omega,\omega)$.  In other words we replicate each detected object, randomizing its $\Omega$, $\omega,$ and $\mathcal{M}$, and weighting inversely by the fraction of randomized objects that are detected.  It is then easy to see that the randomized ensemble has a distribution
\begin{equation}
  p(a,e,i,H | s) \propto \sum_j {\delta(a-a_j) \delta(e-e_j) \delta(i-i_j) \delta(H-H_j)}
  \label{eq:detected}
\end{equation}
and therefore is a precise match to the detected ensemble.  While of course not a realistic model of the underlying eTNO population, it is the simplest way to create a synthetic population that meets the criteria of isotropy and agreement with the distribution of ``uninteresting'' parameters.

To realize the simulated population described by \eqq{eq:parent}, we start by creating 40 million clones of each detected eTNO $j$ for which $\Omega$, $\omega,$ and $\mathcal{M}$ have been redistributed uniformly while retaining $a,e,i,$ and $H$.
We limit the sampling of $\mathcal{M}$ to be uniform between $-15\degr$ and $15\degr$, as all of the detected objects would be too distant and faint to be detected outside this range. This limited sampling translates to a normalization factor of $\frac{30}{360}$ in each $p(s|a_j, e_j, i_j, H_j)$. Since we are only interested in relative detection probabilities, this normalization can be safely ignored. 

For each member of the simulated swarm, we determine all exposures for which the object would be inside a functional DECam CCD and proceed to use the probability $p(m)$ that a point source with magnitude $m$ would be detected in this exposure \citep[see section 2.6 of][]{Bernardinelli2019}.
If $p(m)$ for the simulated object's $m$ is larger than a random unit deviate, this observation is considered a detection of this object.

Once we evaluate all exposures that contain the orbit, we apply the selection criteria used by \citet{Bernardinelli2019} for the \des\ Y4 search: the number of unique nights in which an object was detected must satisfy $\mathtt{NUNIQUE}\ge 6$; the length of the orbital arc must satisfy $\mathtt{ARC}>6$~months; and the shortest arc that remains after eliminating any one night of detections must also satisfy $\mathtt{ARCCUT}>6$~months.

The fraction of all simulated clones of object $j$ that survive these cuts defines the $p(s|a,e,i,H)$ that is in the denominator of \eqq{eq:parent}. Once the simulation is complete, we can calculate the expected $p(\Omega|s)$, $p(\omega|s),$ and $p(\varpi|s)$ of the isotropic population by a histogram of the values for all the clones deemed as detections, weighting inversely by the $p(s)$ values.  Normalizing the histograms to unit integral yields estimates of the probability of detection of an eTNO with angle $\theta \in \{\omega,\Omega,\varpi\}$.
If an object satisfies more than one of our four cases of eTNO definitions, we reuse a single set of clones for all cases, leading to correlations in the small-scale noise of the probabilities for different cases.

\begin{figure}[tb]
	\centering
	\includegraphics[width=0.49\textwidth]{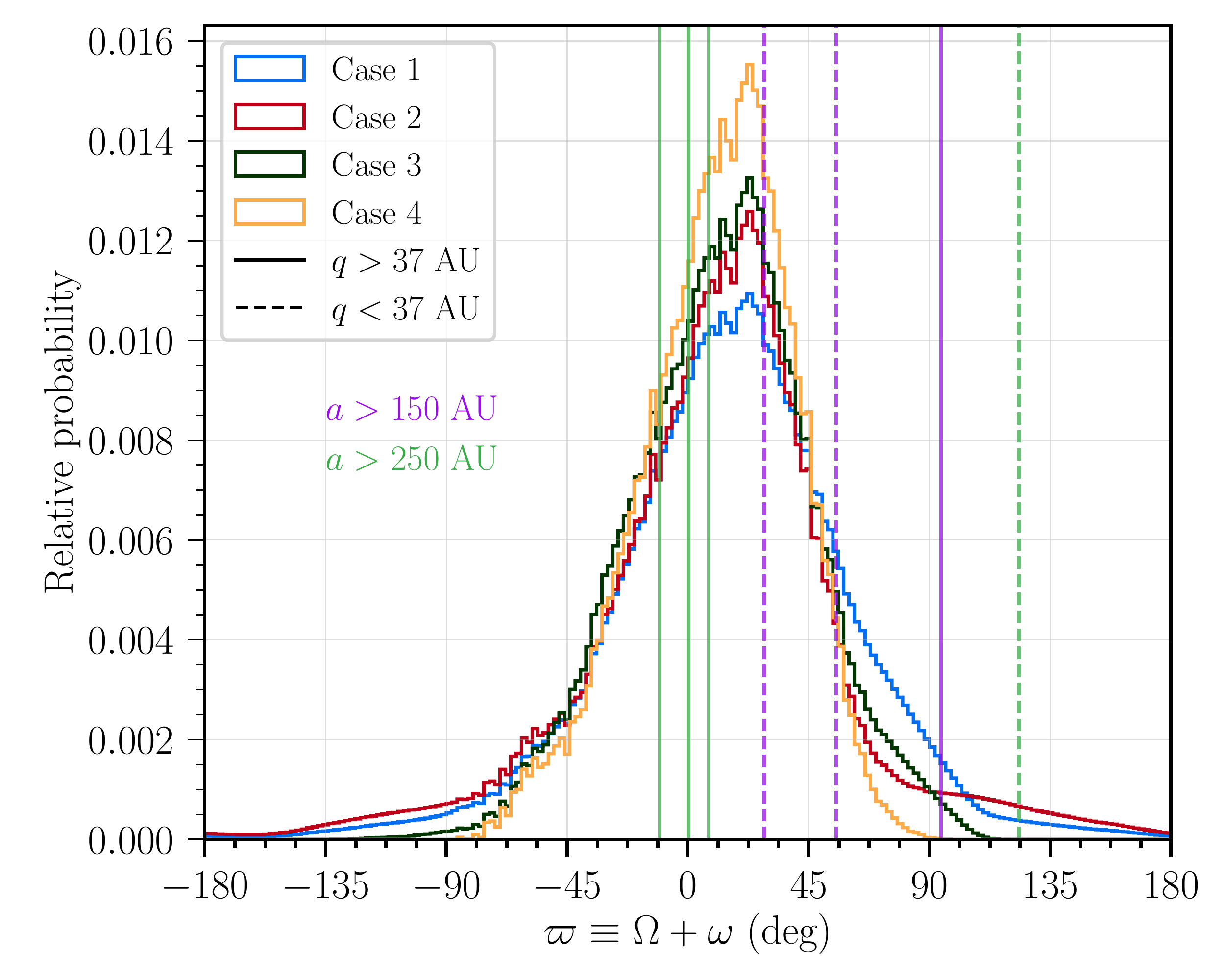}\\
	\includegraphics[width=0.49\textwidth]{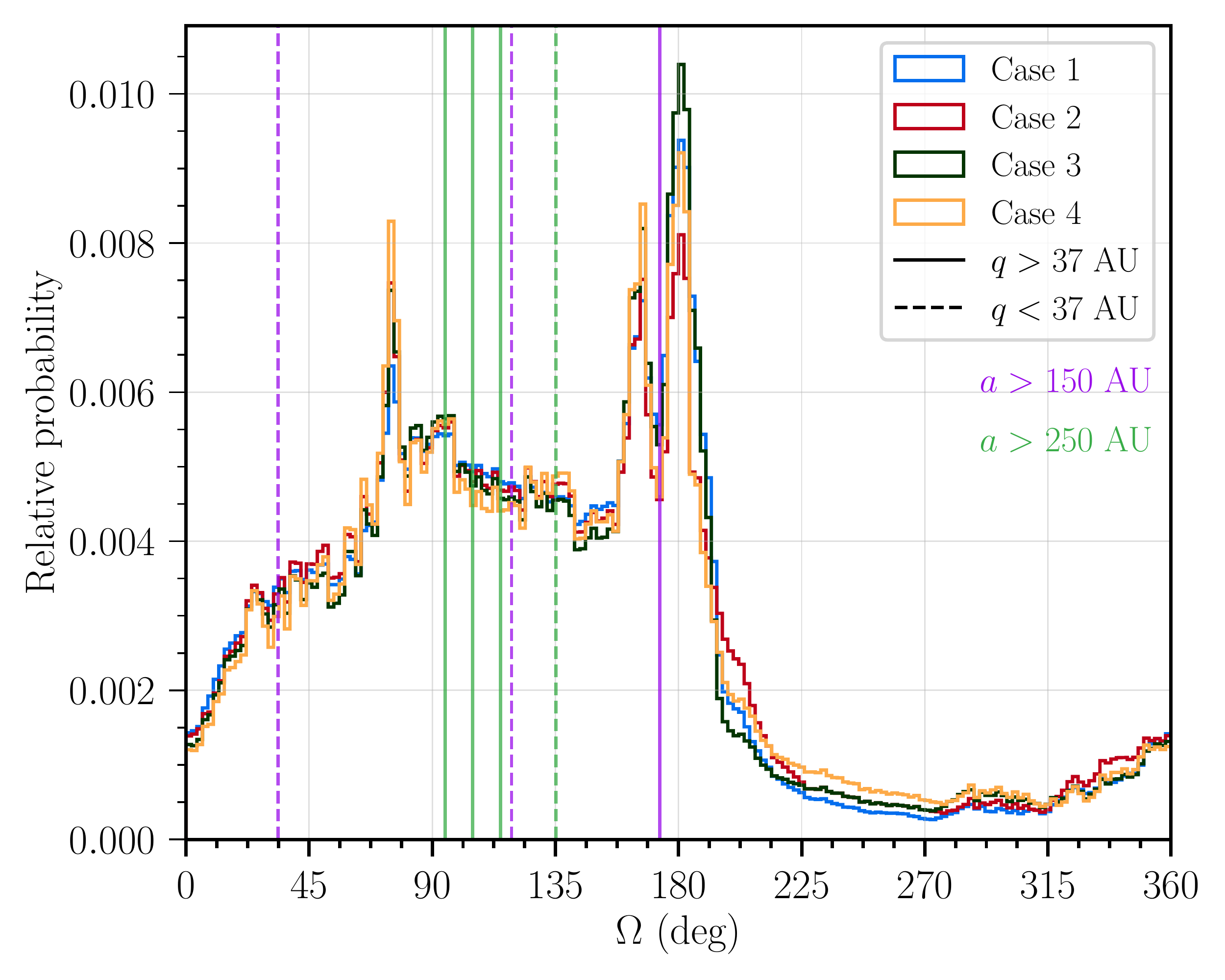}
	\includegraphics[width=0.49\textwidth]{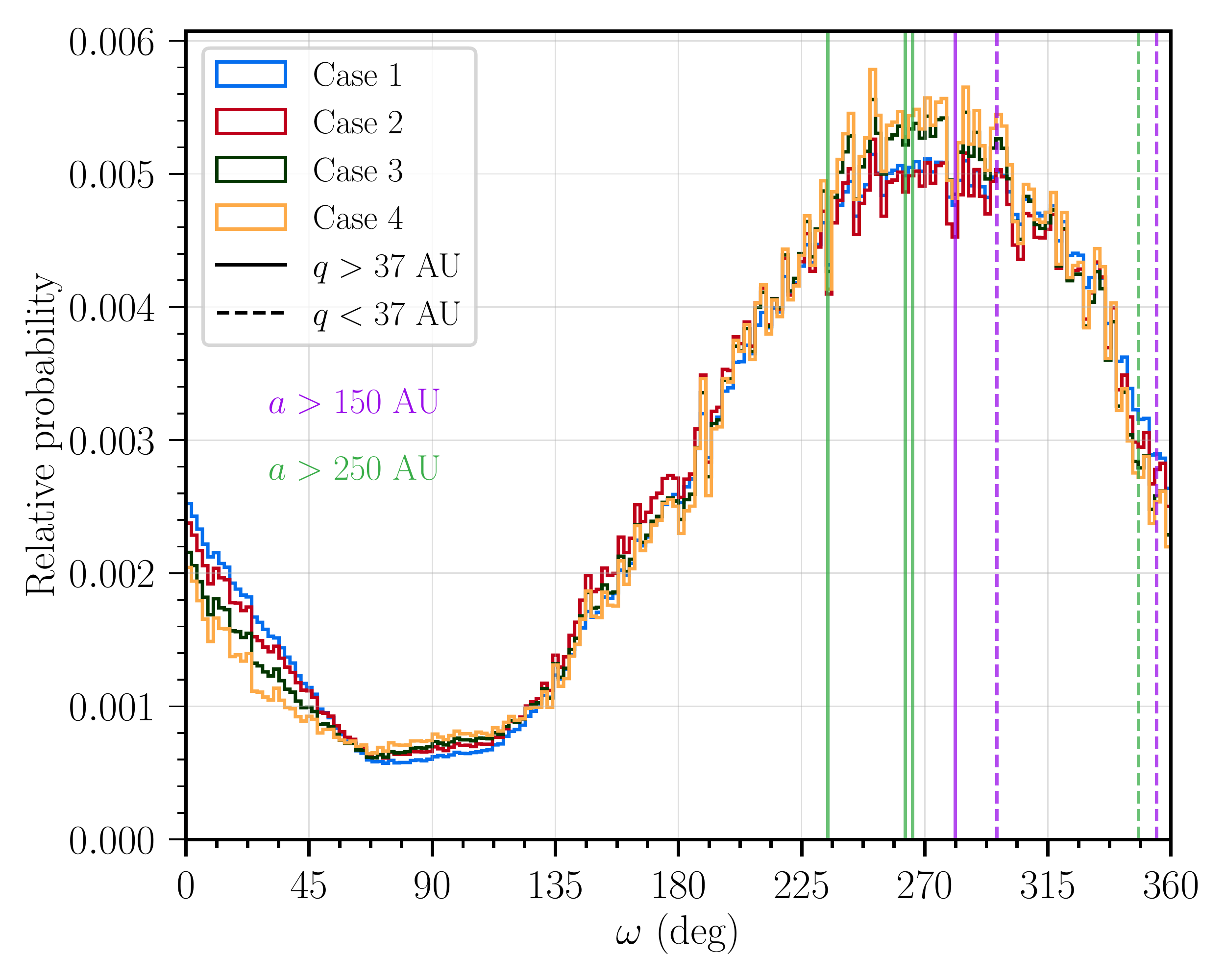}\\

	\caption{Relative probability histograms of $\Omega$, $\omega,$ and $\varpi$ for the \emph{detected} members of a parent population constructed to be intrinsically isotropic in these variables while exactly reproducing the observed $(a,e,i,H)$ distribution of the detected objects.  Histograms normalized to a common integral are shown for each of the four cases of eTNO definitions given in Section \ref{sec:sample}---note that the angular selection functions are very robust to choice of the underlying population. The vertical lines denote the angles at which objects were actually detected in the \des\ Y4 search. The line color denotes the semi-major axis range (purple, $150<a<250$~AU; green, $a > 250$~AU), and the line style denotes the perihelion range (dashed, $30 < q < 37$~AU; solid, $q > 37$~AU). \label{im:angles}}
\end{figure}

Figure \ref{im:angles} shows these angular selection functions for each of the four eTNO definitions and each of the three angles. We note to begin that the selection functions are very similar for all four cases, suggesting that these functions are robust to details of the definition of the $\{a,e,i,H\}$ distribution. The \des\ selection function for longitude of perihelion ($\varpi$) is seen to be quite narrow.  This is not surprising, since the \des\ footprint is confined to a narrow range of ecliptic longitude, and we will have a strong bias toward objects that reach perihelion within the footprint, particularly for the high $e$'s typical of eTNOs.  The strong bias in $\Omega$ seen in Figure~\ref{im:angles} is also easily understood as a consequence of the \des\ footprint being almost entirely in the southern ecliptic hemisphere, in a limited range of ecliptic longitude.

\section{Isotropy tests}
\label{sec:isotropy}

We compare the $p(\theta | s)$ probability distributions derived for an underlying azimuthally isotropic population to the observed distribution of $\theta \in \{\Omega, \omega, \varpi\}$ by applying Kuiper's test \citep{kuiper1960tests}, an extension of the Kolmogorov-Smirnov test that is invariant under cyclic permutations as well as being sensitive both in the median and in the tails of the distribution. For each case of eTNO definition, the significance of this test is measured by computing Kuiper's statistic ${V}_\mathrm{real}$ for the true detected eTNOs to the  $V_{\mathrm{fake}}$ values computed for $10^6$ sets of simulated detections sampled from the isotropized distribution. The $p$-value is the fraction of times $V_\mathrm{fake} > V_{\mathrm{real}}$, i.e. the probability that a Kuiper statistic value as high as the one observed would arise if the angles were drawn from the isotropic population. A test with $p$-value of $0.05$ rejects the null hypothesis with $95\%$ confidence, with lower $p$-values increasing this confidence. We note that this isotropy test is similar to \cite{Shankman2017}'s test on the OSSOS data, in which a population model for eTNOs is built for the null hypothesis. Table~\ref{tb:res} reports the $p$-value of this test for each combination of orbital angle and eTNO definition, for a total of 12 tests.  

The $p$-values for the Kuiper test indicate that the \des\ observations are consistent with being drawn from the isotropic population model, with the possible exception of a low-significance discrepancy ($p\approx0.025$) in the longitude of ascending node ($\Omega$) distribution for cases (2) and (4) at $a>250$~AU. Note that we have performed 12 distinct tests in a small data set, so we cannot claim a significant clustering from a single test at this $p$-value.  Given that the 12 tests are highly correlated, we unfortunately have no straightforward means of determining an overall significance of the ensemble.  If the tests were fully correlated, then of course the chances of observing one at $p\le0.024$ in an isotropic population would be 2.4\%.
If the 12 tests were fully uncorrelated, the chance of having $p\le0.024$ in one or more tests would be $1 - (0.976)^{12} = 25 \%$; these can be considered lower and upper bounds on the overall significance of anisotropy.
The $p$-values of the ensembles remain very sensitive to small changes in the eTNO definition, due to the small number of detections, which counsels further caution in assigning significance to the appearance of $p\approx0.025$ values in our ensemble of tests.  For example, adopting an eTNO definition of $a>230$~AU \citep[following][]{Brown2017,Brown2019}, yields $p$-values for the $\{\omega,\Omega,\varpi\}$ distributions of $\{0.468, 0.006, 0.532\}$, respectively.  While the nodal clustering becomes apparently stronger, this is not the only variable in which \citet{Brown2019} find a signal for their sample, and there is no evidence for clustering in $\omega$ or $\varpi$, the variables in which they reported the strongest TNO alignments.
Perhaps the most conservative approach would be to examine only the test for $\varpi,$ the variable previously found to have the strongest clustering, using Case (4), which isolates the objects most sensitive to the dynamical effect of Planet 9 and least influenced by Neptune.  For this single test, $p=0.11,$ meaning the null hypothesis of isotropy is rejectable with only $89\%$ confidence.

One other statistic that we can use to judge the agreement between the observed and isotropized populations is the overall likelihood of the observed values of orbital angle $\theta \in \{ \Omega, \omega, \varpi\}$:
\begin{equation}
  \mathcal{L} \equiv p\left(\{\theta_j\}\right) = \prod_j p(\theta_j | s),
  \label{eq:prob}
\end{equation}
taking the probability densities $p(\theta | s)$ directly from the simulation-derived histograms in Figure~\ref{im:angles}.
While the ensemble likelihoods $\mathcal{L}$ are not themselves readily interpretable, we can produce an expected cumulative distribution function for $\mathcal{L}$ under the null hypothesis (isotropy) by calculating it for a large number of sets of ``detections'' drawn at random from the simulated population.  We denote by $f$ the fraction of sets of simulated isotropic detections that yield $\mathcal{L}$ lower than that for the true detected objects. The $f$-test is more sensitive to individual objects being detected at the tails of the isotropic distribution, but unlike the Kuiper test it does not consider the collective distribution.  For example the $f$ test would not register an abnormality if all the detections were on one side of a symmetric distribution. So the tests can be seen as complimentary and should not be expected to yeld similar significance.

The $f$-values for each combination of orbital angle $\theta \in \{\Omega, \omega, \varpi\}$ and eTNO definition are also listed in Table~\ref{tb:res}. For $\varpi$, all $f$-values are in the $20\sim30$\% range, and in the $50\sim60$\% range for $\Omega$, so the measured angles are not particularly likely or unlikely given the survey's selection functions. All eTNO definitions present a somewhat high $f$ ($> 90\%$) for $\omega$, meaning that these detections are among the most likely outcomes possible given the isotropized distribution. This is not a surprise, since visual inspection of the $\omega$ selection functions (Figure \ref{im:angles}) shows that all objects are in the region of highest probability. In sum, the $\mathcal{L}$ statistics are fully compatible with the null hypothesis of isotropy.

\begin{deluxetable}{|c|cc|cc|cc|}
	\tablecaption{$p$-values derived using Kuiper's test applied to the four distinct eTNO definitions (Section \ref{sec:sample}) studied here measuring how likely it is that the measured objects come from a uniform underlying distribution. The $f$ values represent the fraction of simulated isotropic detections that yield a likelihood $\mathcal{L}$ lower than the one for the true objects. Lower $f$ or $p$ values represent more significant deviations from isotropy.\label{tb:res}}
	\tablehead{Case & $p(\varpi)$ & $f(\varpi)$ & $p(\Omega)$ & $f(\Omega)$ & $p(\omega)$ & $f(\omega)$}
	\startdata
		Case 1 & 0.933 & 0.235 & 0.180 & 0.595 & 0.393 & 0.960 \\
		Case 2 & 0.313 & 0.282 & 0.028 & 0.525 & 0.326 & 0.938 \\
		Case 3 & 0.361 & 0.192 & 0.211 & 0.628 & 0.053 & 0.973 \\
		Case 4 & 0.109 & 0.300 & 0.024 & 0.498 & 0.072 & 0.933 \\
	\enddata
\end{deluxetable}

\section{Alternative analysis}
\label{sec:stephanie}
A distinct analysis of the isotropy of \des\ extreme-TNO detections is reported in full by \citet{Hamilton2019}, reaching the same conclusions as presented above, namely that the \des\ data do not by themselves offer strong evidence of alignments in the outermost known solar system.
We highlight the major ways in which the \citet{Hamilton2019} analysis differs from that presented above---details can be found in the publication.
\begin{itemize}
\item Single-night transients were discovered using difference imaging \citep{Kessler2015,Herner2020}, rather than the catalog-level comparisons of \citet{Bernardinelli2019}.
\item The difference imaging was executed on a subset of the first three years of \des\ imaging, rather than on the full four-year data reported herein.
\item The alternative analysis includes TNOs discovered in the \des\ supernova-search fields, whereas the Y4 analysis herein does not.
\item Different software and algorithms were used to link TNOs from the collection of detected transients.
\item The detection completeness of individual exposures for point sources was determined by measuring the signal-to-noise ratio of sources of fixed, bright magnitude injected directly into the images, and calibrating this $S/N$ level into a point-source completeness threshold \citep{Kessler2015}.  The method of \citet{Bernardinelli2019} is to determine detection efficiency vs.\ magnitude using faint stars in the fields.
\item The alternative analysis creates expected distributions for $\Omega, \varpi,$ and $\omega$ using a null hypothesis positing a chosen smooth distribution of sources in the space of $\{a,e,i,H\}$ with isotropy in $\{\Omega,\omega\}$, as opposed to this paper's technique of building the null-hypothesis population from isotropized copies of the discovered objects.
\end{itemize}
This difference-imaging search yields a sample of 4 TNOs meeting a definition of ``extreme'' as $a>250$~AU, $q>30$~AU, the
same as case (2) above---although these are not the same 4 objects as in the case (2) analysis: 2013 RF$_{98}$ was discovered in the \des\ supernova-search fields, while 2014 WB$_{556}$ had not been discovered. 

Figures~5.1 of \citet{Hamilton2019} present the null-hypothesis and the observed distributions of $\Omega,\omega,$ and $\varpi$ in analogy with Figures~\ref{im:angles} above, and look very similar despite very different implementations of the processing steps.  The Kuiper test statistic for departures from isotropy in $\Omega,\omega,$ and $\varpi$ are found to be exceeded by 8\%, 24\%, and 43\% of the null-hypothesis distributions, respectively (see Figure~5.3).  This leads to the same conclusions as the corresponding values of 3\%, 32\%, and 33\% for Case 2 in Table~\ref{tb:res}.

\section{Conclusion}
\label{sec:conclusions}

We succeeded with little difficulty in creating an isotropic population model for eTNOs that matches the \des\ observations. The populations at $a>250$~AU are only marginally compatible with isotropy in $\Omega$ ($p\approx0.025$, $f \approx 0.5$), but this discrepancy is not strong enough to falsify the isotropy hypothesis given the small samples and multiple variables that we test. Similar to \citet{Shankman2017}'s analysis of the well-characterized OSSOS data, our analysis of the \des\ data does not present evidence of the Planet 9 hypothesis. 
We note that the consistency with an isotropic model does not falsify the Planet 9 hypothesis.  Falsification would require that one show that \emph{all} population models under this hypothesis are \emph{inconsistent} with the data.
The \des\ selection function is narrow in $\varpi$, reducing our sensitivity to true anisotropies. On the other hand, with a larger sample any $\varpi$ distribution that is not constant across our limited window would eventually be detectable.
When the full six years of \des\ observations are analyzed, the geometry of the selection functions should not change much, but the final catalog is expected to yield detections 0.5 magnitudes deeper, likely
increasing the total number of eTNOs in our sample.

\acknowledgments

The authors thank Rodney Gomes and Hsing Wen Lin, as well as the anonymous referees, for reviewing this manuscript and providing feedback which led to the improvement of this work . University of Pennsylvania authors have been supported in this work by grants AST-1515804 and AST-1615555 from the National Science Foundation, and grant DE-SC0007901 from the Department of Energy. Work at University of Michigan is supported by the National Aeronautics and Space Administration under Grant No. NNX17AF21G issued through the SSO Planetary Astronomy Program and  NSF Graduate Research Fellowship Grant No. DGE 1256260.

Funding for the DES Projects has been provided by the U.S. Department of Energy, the U.S. National Science Foundation, the Ministry of Science and Education of Spain,
the Science and Technology Facilities Council of the United Kingdom, the Higher Education Funding Council for England, the National Center for Supercomputing
Applications at the University of Illinois at Urbana-Champaign, the Kavli Institute of Cosmological Physics at the University of Chicago,
the Center for Cosmology and Astro-Particle Physics at the Ohio State University,
the Mitchell Institute for Fundamental Physics and Astronomy at Texas A\&M University, Financiadora de Estudos e Projetos,
Funda{\c c}{\~a}o Carlos Chagas Filho de Amparo {\`a} Pesquisa do Estado do Rio de Janeiro, Conselho Nacional de Desenvolvimento Cient{\'i}fico e Tecnol{\'o}gico and
the Minist{\'e}rio da Ci{\^e}ncia, Tecnologia e Inova{\c c}{\~a}o, the Deutsche Forschungsgemeinschaft and the Collaborating Institutions in the Dark Energy Survey.

The Collaborating Institutions are Argonne National Laboratory, the University of California at Santa Cruz, the University of Cambridge, Centro de Investigaciones Energ{\'e}ticas,
Medioambientales y Tecnol{\'o}gicas-Madrid, the University of Chicago, University College London, the DES-Brazil Consortium, the University of Edinburgh,
the Eidgen{\"o}ssische Technische Hochschule (ETH) Z{\"u}rich,
Fermi National Accelerator Laboratory, the University of Illinois at Urbana-Champaign, the Institut de Ci{\`e}ncies de l'Espai (IEEC/CSIC),
the Institut de F{\'i}sica d'Altes Energies, Lawrence Berkeley National Laboratory, the Ludwig-Maximilians Universit{\"a}t M{\"u}nchen and the associated Excellence Cluster Universe,
the University of Michigan, the National Optical Astronomy Observatory, the University of Nottingham, The Ohio State University, the University of Pennsylvania, the University of Portsmouth,
SLAC National Accelerator Laboratory, Stanford University, the University of Sussex, Texas A\&M University, and the OzDES Membership Consortium.

Based in part on observations at Cerro Tololo Inter-American Observatory, National Optical-Infrared Astronomy Observatory, which is operated by the Association of 
Universities for Research in Astronomy (AURA) under a cooperative agreement with the National Science Foundation.

The DES data management system is supported by the National Science Foundation under Grant Numbers AST-1138766 and AST-1536171.
The DES participants from Spanish institutions are partially supported by MINECO under grants AYA2015-71825, ESP2015-66861, FPA2015-68048, SEV-2016-0588, SEV-2016-0597, and MDM-2015-0509,
some of which include ERDF funds from the European Union. IFAE is partially funded by the CERCA program of the Generalitat de Catalunya.
Research leading to these results has received funding from the European Research
Council under the European Union's Seventh Framework Program (FP7/2007-2013) including ERC grant agreements 240672, 291329, and 306478.
We acknowledge support from the Australian Research Council Centre of Excellence for All-sky Astrophysics (CAASTRO), through project number CE110001020, and the Brazilian Instituto Nacional de Ci\^encia
e Tecnologia (INCT) e-Universe (CNPq grant 465376/2014-2).

This manuscript has been authored by Fermi Research Alliance, LLC under Contract No. DE-AC02-07CH11359 with the U.S. Department of Energy, Office of Science, Office of High Energy Physics. The United States Government retains and the publisher, by accepting the article for publication, acknowledges that the United States Government retains a non-exclusive, paid-up, irrevocable, world-wide license to publish or reproduce the published form of this manuscript, or allow others to do so, for United States Government purposes.

\bibliography{references}
\bibliographystyle{aasjournal}

\end{document}